\documentclass[aps,prc,preprint,amsmath,amssymb,showpacs,preprintnumbers,superscriptaddress]{revtex4-1}
\usepackage{graphicx}
\usepackage{dcolumn}
\usepackage{bm}

\usepackage{hyperref}

\begin{document}


\title{Impact of pairing correlations on the orientation of the nuclear spin}

\author{P. W. Zhao 
}
\email{pwzhao@pku.edu.cn}
\affiliation{Physics Division, Argonne National Laboratory, Argonne, Illinois 60439, USA}
\affiliation{State Key Laboratory of Nuclear Physics and Technology, School of Physics, 
Peking University, Beijing 100871, China}

\author{S. Q. Zhang 
}
\affiliation{State Key Laboratory of Nuclear Physics and Technology, School of Physics, 
Peking University, Beijing 100871, China}

\author{J. Meng 
}
\affiliation{State Key Laboratory of Nuclear Physics and Technology, School of Physics, 
Peking University, Beijing 100871, China}
\affiliation{School of Physics and Nuclear Energy Engineering, Beihang University, Beijing 100191, China}
\affiliation{Department of Physics, University of Stellenbosch, Stellenbosch 7602, South Africa}


\begin{abstract}
For the first time, the tilted axis cranking covariant density functional theory with pairing correlations has been formulated and implemented in a fully self-consistent and microscopic way to investigate the evolution of the spin axis and the pairing effects in rotating triaxial nuclei. 
The measured energy spectrum and transition probabilities for the $^{135}$Nd yrast band are reproduced well without any ad hoc renormalization factors when pairing effects are taken into account.
A transition from collective to chiral rotation has been demonstrated. 
It is found that pairing correlations introduce additional admixtures in the single-particle orbitals, and, thus, influence the structure of tilted axis rotating nuclei by reducing the magnitude of the proton and neutron angular momenta while merging their direction. 
\end{abstract}


\maketitle



\section{Introduction}
Similar to rotational bands observed in molecules, the most common collective excitation in nuclei corresponds to a rotation about the principal axis of the density distribution with the larger moment of inertia. The rotation is collective since a large fraction of the angular momentum is generated through small contributions from many nucleons. It is well known that nuclear deformation and superfluidity play vital roles in generating  angular momentum~\cite{Bohr1975}. 
The substantial deformation of the overall density distribution specifies the orientation of a nucleus and, thus, the rotational degree of freedom. Meanwhile, its superfluid behavior is required by the fact that the observed collective moment of inertia is usually much smaller than the rigid-body estimate. 

Unlike molecules, nuclei can rotate about an axis tilted with respect to the principal axes of the density distribution~\cite{Frauendorf2001Rev.Mod.Phys.463} due to the fact that a nucleus is composed of nucleons carrying a quantized amount of angular momentum. This is the so-called tilted axis rotation which was first proposed within the mean-field tilted axis cranking (TAC) approach~\cite{Frauendorf1993Nucl.Phys.A259,Frauendorf2000Nucl.Phys.A115}.
The tilt of the rotational axis is closely related to the interplay between collective and single-nucleon motions. Therefore, the two elements, deformation and superfluidity, are not only crucial in generating the magnitude of the spin  but also its orientation.  

A variety of discrete symmetries for rotating nuclei can be obtained by combining the deformation and the spin  orientation, and this gives rise to a variety of new phenomena. Among the latter figure, the magnetic and antimagnetic rotation in nearly spherical nuclei~\cite{Frauendorf2001Rev.Mod.Phys.463,Clark2000Annu.Rev.Nucl.Part.Sci.1,Hubel2005Prog.Part.Nucl.Phys.1}, the high-$K$ bands giving rise to $K$ isomerism~\cite{Walker1999Nature35} in axially deformed nuclei, etc. 
For triaxial nuclei, specifically, the rotational axis may lie outside the three principal planes of the ellipsoidal shape. This forms the so-called aplanar rotation and causes nuclear chirality~\cite{Frauendorf1997Nucl.Phys.A131}, a mode that has attracted significant attention due to its importance on the subatomic physics scale ~\cite{Frauendorf2001Rev.Mod.Phys.463,Meng2010JPhysG.37.64025}. 
For axially deformed nuclei, the rotational axis always lies in the plane defined by the symmetry axis and the one perpendicular to it. In contrast, planar triaxial solutions can be found in the three distinct principal planes. 
This also leads to many new interesting modes such as chiral vibration~\cite{Mukhopadhyay2007Phys.Rev.Lett.172501,Qi2009Phys.Lett.B175} and transverse wobbling~\cite{Frauendorf2014Phys.Rev.C14322,Chen2014Phys.Rev.C44306,Matta2015Phys.Rev.Lett.82501}. 
Furthermore, a rotating triaxial nucleus allows more degrees of freedom for the evolution of the spin axis. This makes it all the more interesting to investigate, for example, a transition from planar to aplanar rotation, i.e., from chiral vibration to static chirality~\cite{Zhu2003Phys.Rev.Lett.132501,Mukhopadhyay2007Phys.Rev.Lett.172501,Qi2009Phys.Lett.B175}, and a transition from one planar rotation to another; this has not been studied in any detail so far.   

In contrast to the well-known impact of pairing on principal axis rotation~\cite{Ring1980}, its influence on a tilted axis rotor is still far from being understood. 
In particular, because it affects both the collective and valence nucleon motions, it should be expected that pairing correlations may reorient the spin axis in tilted rotation. So far, most calculations are based on single-particle potentials combined with the pairing plus quadrupole-quadrupole model~\cite{Frauendorf2000Nucl.Phys.A115}.
Therefore, self-consistent methods based on more realistic two-body interactions are required for a more fundamental investigation, including all important effects such as core polarization and nuclear currents~\cite{Olbratowski2004Phys.Rev.Lett.52501,Zhao2011Phys.Rev.Lett.122501,Meng2013FrontiersofPhysics55}. 
Such calculations are more challenging, but are feasible in the framework of both relativistic~\cite{Madokoro2000Phys.Rev.C61301,Peng2008Phys.Rev.C24313,Zhao2011Phys.Lett.B181} and nonrelativistic~\cite{Olbratowski2002ActaPhys.Pol.B389,Olbratowski2004Phys.Rev.Lett.52501,Olbratowski2006Phys.Rev.C54308} density functional theories (DFTs). 
However, pairing correlations have not been taken into account in any of these studies. For many years, the computed bandhead energies and magnetic dipole transition probabilities $B(M1)$ had to be renormalized by ad hoc factors to reproduce the data~\cite{Madokoro2000Phys.Rev.C61301,Zhao2011Phys.Lett.B181,Yu2012Phys.Rev.C24318}, leading to a long-standing question about whether the inclusion of pairing correlations would improve the agreement between data and calculations. 

The focus of the present research is two-fold: 1) the transition of the spin axis from one principal plane to another in a triaxial nucleus; 2) the impact of pairing correlations in a tilted axis rotor. 
This article presents the first tilted axis cranking covariant DFT with pairing correlations. The spin axis evolution and the impact of pairing for a rotating triaxial nucleus have been investigated in a fully self-consistent microscopic way for the first time. 
Covariant DFT~\cite{Ring1996Prog.Part.Nucl.Phys.193,Vretenar2005Phys.Rep.101,Meng2006Prog.Part.Nucl.Phys.470} consistently treats the spin degrees of freedom, includes the complex interplay between the large Lorentz scalar and vector self-energies induced at the QCD level~\cite{Cohen1992Phys.Rev.C1881}. Moreover, the nuclear currents 
which are essential for rotating nuclei, are provided naturally from the spatial parts of the vector self-energies. 

\section{Theoretical Framework}
For a unified and self-consistent treatment of the mean fields and pairing correlations, one has to solve the fully relativistic Hartree-Bogoliubov (RHB) problem~\cite{Niksic2014Comput.Phys.Commun.} in the framework of superfluid covariant DFT. 
The RHB model contains two average potentials: the mean fields $S(\bm{r})$ and $V^\mu(\bm{r})$ which include all the long range particle-hole (\emph{ph}) correlations, and a pairing field $\Delta (\bm{r})$ which sums up the particle-particle (\emph{pp}) correlations. 
In the TAC model, these potentials are deformed and the calculations are carried out in the intrinsic frame rotating with a constant angular velocity vector $\bm{\omega}$ pointing in a direction which is not parallel to one of the principal axes of the density distribution:
\begin{equation}\label{RHBequation}
\left(
  \begin{array}{cc}
   h-\bm{\omega}\cdot\hat{\bm{J}} & \Delta \\
    -{\Delta}^\ast & -h^\ast+\bm{\omega}\cdot\hat{\bm{J}}^\ast \\
  \end{array}
\right)
\left(
   \begin{array}{c}
     U_k \\
     V_k \\
   \end{array}
 \right)
=E_k\left(
   \begin{array}{c}
     U_k \\
     V_k \\
   \end{array}
 \right).
\end{equation}
Here $h=h_D-\lambda$ is the single-nucleon Dirac Hamiltonian 
\begin{equation}
   h_D = \bm{\alpha}\cdot(\bm{p}-\bm{V})+\beta(m+S)+V
 \end{equation}
minus the chemical potential $\lambda$, and $\hat{\bm{J}}$ is the total angular momentum of the nucleon spinors. 
$U_k$ and $V_k$ are the quasiparticle Dirac spinors and $E_k$ denotes the quasiparticle (qp) energies. 
The mean fields $S$ and $V^\mu$ as well as the pairing field $\Delta$ are connected in a self-consistent way to the densities and currents as well as to the pairing tensor distributions.
The iterative solution of these equations yields expectation values of the angular momentum, total energies, quadrupole moments, transition probabilities~\cite{Zhao2012Phys.Rev.C54310}, etc.
The magnitude of the angular velocity $\bm{\omega}$ is connected to the angular momentum quantum number $I$ by the semiclassical relation $\langle\hat{\bm{J}}\rangle\cdot\langle\hat{\bm{J}}\rangle=I(I+1)$.

The observed yrast band in the odd-$A$ nucleus $^{135}$Nd~\cite{Zhu2001Phys.Rev.C41302} is investigated in the present work. The ground band is associated with the $\nu h_{11/2}$ one quasi-neutron configuration~\cite{Beck1987Phys.Rev.Lett.2182}. 
However, above $I=29/2\hbar$, this configuration is further coupled to two aligned $h_{11/2}$ protons, thereby resulting in the 3-qp configuration $\nu h_{11/2}\pi h_{11/2}^2$. 
This 3-qp band and its partner have been interpreted as a pair of chiral bands~\cite{Zhu2003Phys.Rev.Lett.132501,Qi2009Phys.Lett.B175}, a property supported by lifetime measurements~\cite{Mukhopadhyay2007Phys.Rev.Lett.172501}. 
The present self-consistent investigation includes both the 1-qp and 3-qp bands, and the evolution of the rotational axis will be analyzed. 
The point-coupling energy density functional PC-PK1~\cite{Zhao2010Phys.Rev.C54319} is adopted in the \emph{ph} channel, and a monopole pairing force with constant strength, for neutrons $G_n=0.12$ MeV~fm$^3$ and for protons $G_p=0.13$ MeV~fm$^3$, determined from the odd-even mass differences is used in the \emph{pp} channel. The calculations are free of additional parameters. 

In this work, we allow only rotations around an axis in the ($x,z$) plane. Equation~(\ref{RHBequation}) is solved in a three-dimensional Cartesian harmonic oscillator basis~\cite{Koepf1989Nucl.Phys.A61} with $N=$10 major shells. 
Parity is the only good quantum number, and the space of the Hamiltonian matrix is, thus,  twice as large as for the principal axis cranking RHB theory~\cite{Afanasjev1999Phys.Rev.C51303}. Therefore, parallel computations have been implemented to reduce the required computational time.
Another difficulty is that one has to trace and block the right qp orbitals to keep the multi-qp configuration unchanged while solving Eq.~(\ref{RHBequation}) iteratively with different $\lambda$ and $\bm{\omega}$ values.
To achieve this, we first define for each single qp state the normalized wave functions
$\psi^u_k = {U_k}/{\sqrt{(U_k)^T U_k}}$ and $\psi^v_k ={V_k}/{\sqrt{(V_k)^TV_k}}$. 
Then, we search at each iteration for the largest overlap 
$$O_1 = \langle \psi^u_k|\psi^u_{k'_1}\rangle+ \langle\psi^v_k|\psi^v_{k'_1} \rangle,\, 
O_2=\langle \psi^u_k|\psi^v_{k'_2} \rangle + \langle \psi^v_k|\psi^u_{k'_2} \rangle,$$
where $k'_1$ ($k'_2$) are determined by running over all the single qp states obtained in the previous step while maximizing the overlap.
Finally, the state $k$ would be blocked only if $O_1$ is larger (smaller) than $O_2$ and the state $k'_1$ ($k'_2$) was blocked (unblocked).

\section{Results and discussion}
\begin{figure}[!htbp]
\centering
\includegraphics[width=8cm]{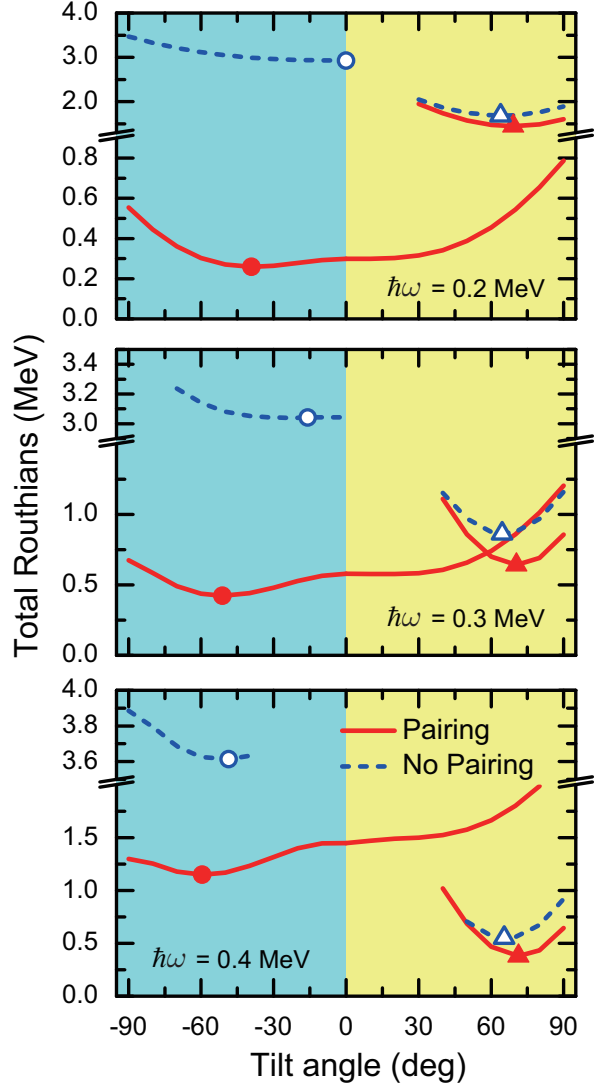}
\caption{(Color online) Total Routhians for the 1-qp and 3-qp configurations as functions of the tilt angle $\theta$ at the rotational frequencies $\hbar\omega= 0.2$ (top), 0.3 (center), 0.4 (bottom) MeV. The circles and triangles represent the local minimum points for the 1-qp and 3-qp configurations, respectively.}
\label{fig1}
\end{figure}

The present self-consistent calculation for $^{135}$Nd leads automatically for the bandhead to a state with considerable triaxial deformation ($\gamma\sim22^\circ$) associated with the 1-qp configuration $\nu h_{11/2}$. 
Along the band, the tilt angle $\theta$ of the rotational axis is determined in a self-consistent way by minimizing the total Routhians (Fig.~\ref{fig1}). 
Here, the tilt angle is defined as the angle between the rotational and the long axis, and the positive (negative) value denotes a tilt towards the short (intermediate) axis. 
Figure~\ref{fig1} clearly indicates that the 1-qp state always favors rotating along an axis in the long-intermediate ($l$-$i$) plane, and the rotation axis tilts more and more appreciably toward the $i$ axis with increasing frequency. 
Moreover, the 3-qp state, rotating along an axis in the long-short ($l$-$s$) plane, becomes lower in energy than the 1-qp configuration at high frequency.
This is consistent with the fact that the observed 3-qp band becomes yrast at high angular momentum. 
The pairing effects for the 3-qp states are suppressed significantly by the two aligned $h_{11/2}$ protons. In contrast, the 1-qp states are lifted notably in energy by excluding the pairing correlations, and, thus, they are always located much higher than the corresponding 3-qp states. 
Therefore, it is concluded that one can determine the band crossing frequency more accurately by including the pairing correlations.

\begin{figure}[!htbp]
\centering
\includegraphics[width=8cm]{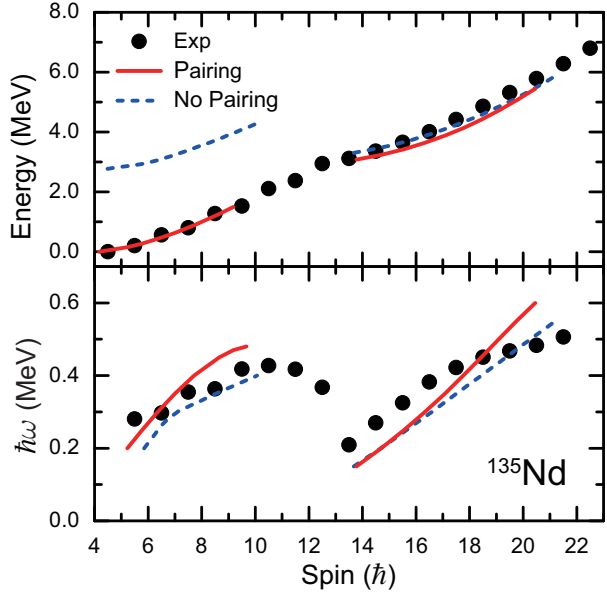}
\caption{(Color online) Rotational  energy (upper panel) and rotational frequency (lower panel) as functions of the angular momentum in comparison with the data of Ref.~\cite{Zhu2003Phys.Rev.Lett.132501} (solid dots). Here, the excitation energies are the relative energy differences with respect to the ground state. 
}
\label{fig2}
\end{figure}

In addition to determining the tilt angle, the calculated energy spectrum and the angular momenta could be calculated as well: these are compared with the data~\cite{Zhu2003Phys.Rev.Lett.132501} in Fig.~\ref{fig2} with the upper panel indicating that the experimental rotational excitation energies for both the 1-qp (lower spin part) and 3-qp bands (higher spin part) are reproduced well by the present self-consistent calculations with pairing. 
In particular, the inclusion of pairing leads to the correct energy difference between the 1-qp and 3-qp configurations. 
Therefore, the need for an artificial renormalization of the bandhead energies mentioned above has been eliminated by including pairing. 

In the lower panel of Fig.~\ref{fig2}, it is seen that the calculated total angular momenta also agree well with the data. 
A backbending happens in the region $I=10$-$14\hbar$, where the angular momentum increases while the rotational frequency drops drastically. 
It is wellknown that such a phenomenon is beyond the scope of a cranking calculation~\cite{Hamamoto1976Nucl.Phys.A15}, and, consequently, the calculated results are omitted.
It should be expected that the moment of inertia $I/\omega$ is reduced by pairing in the present tilted axis calculations. 
However, it is surprising that the reduction for the 3-qp band is only visible at high frequency, where the pairing effects tend to be hindered by the Coriolis term. 
Indeed, here the proton pairing gaps are very small due to the two quasi-protons, while the neutron ones decreases from 1.25 MeV ($\hbar\omega=0.2$ MeV) to 1.14 MeV ($\hbar\omega= 0.6$ MeV). 

\begin{figure*}[!htbp]
\centering
\includegraphics[width=14cm]{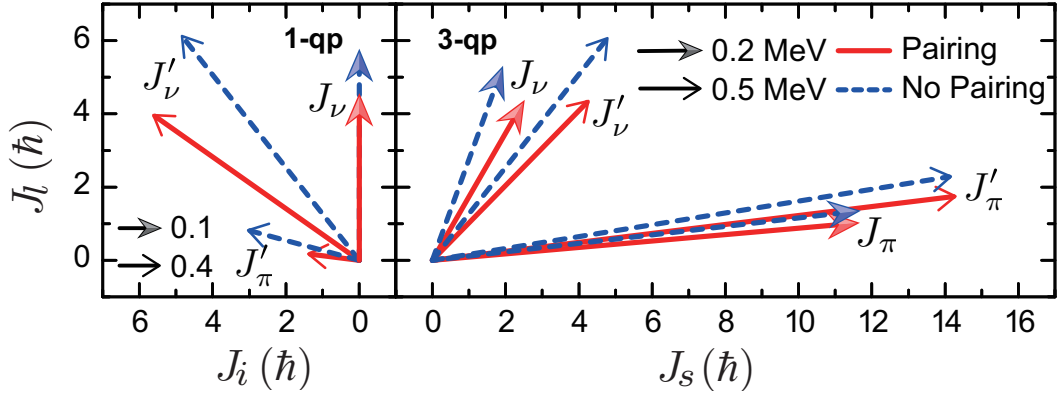}
\caption{(Color online) Neutron $\bm{J}_\nu$ and proton $\bm{J}_\pi$ angular momentum vectors for both the 1-qp (left) and 3-qp (right) bands. Different types of arrows denote the results at different rotational frequencies. Note that $\bm{J}_\pi$ for the 1-qp band is negligible at $\hbar\omega=0.1$ MeV.}
\label{fig3}
\end{figure*}

To understand this distinctive feature, the neutron and proton angular momentum vectors are provided in Fig.~\ref{fig3} for both the 1-qp and 3-qp bands. 
For the former band, the neutron angular momentum aligns along the $l$ axis and the proton one essentially vanishes at $\hbar\omega= 0.1$ MeV.
Along the band, the neutron angular momentum keeps its projection on the $l$ axis nearly constant, which reflects the 
contribution of the one unpaired neutron hole in the $h_{11/2}$ shell, but at the same time the proton angular momentum increases remarkably along the $i$ axis due to a coherent collective rotation of many nucleons. 

For the 3-qp band, the neutron angular momentum aligns mainly along the $l$ axis due to the $h_{11/2}$ hole, while the proton one aligns mainly along the $s$ axis, due to the two $h_{11/2}$ particles involved. 
As the frequency increases, the neutron and proton angular momenta align toward each other and generate larger total angular momentum with the direction nearly unchanged. 
This situation is reminiscent of the ``shears mechanism'' in a magnetic rotation band. 
However, due to the considerable triaxiality ($\gamma\sim22^\circ$) involved, it has been shown in Ref.~\cite{Mukhopadhyay2007Phys.Rev.Lett.172501} that a chiral vibration, resulting from rapid conversion between the left-handed and right-handed configurations, has been realized in the 3-qp band. Therefore, a transition from collective to chiral rotation is observed here.  

In contrast to principal axis rotation, pairing effects on the angular momentum here result in two competing effects. 
On the one hand, the magnitudes of both proton and neutron angular momenta are reduced by pairing. 
On the other, however, pairing tends to reduce the angle between the proton and neutron angular momenta (by up to 20\%) and in this way increases the total spin. 
Therefore, the impact of pairing on the total angular momentum can be, in some cases, counteracted as is the case for the lower spin part of the 3-qp band in Fig.~\ref{fig2}. All in all, pairing introduces superfluidity in rotational states and, here, it expedites the closing of the proton and neutron angular momenta. 

\begin{figure}[!htbp]
\centering
\includegraphics[width=7cm]{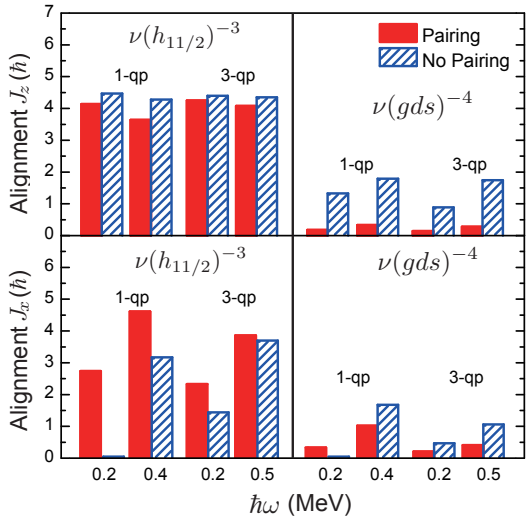}
\caption{(Color online) Angular momentum alignments for the 1-qp and 3-qp configurations on the $z$ axis (upper panels) and $x$ axis (lower panels) of the neutron holes in the $h_{11/2}$ shell (left) and the $(g_{7/2}d_{5/2}d_{3/2}s_{1/2})$ shell (right). }
\label{fig4}
\end{figure}

To trace the microscopic reason for the pairing effects, it would be quite helpful to use the Bogoliubov transformation to transform the qp basis to the canonical (particle) basis~\cite{Ring1980}, as in a microscopic picture, the angular momentum comes from all the individual particles. Since the pairing effects on the protons are significantly blocked in the 3-qp band, we show in Fig.~\ref{fig4} only the angular momentum contributions of the  neutron holes while noting that the proton ones lead to a similar conclusion.  

For the 75 neutrons in $^{135}$Nd, there are contributions from only the seven neutron holes with respect to the closed $N=82$ shell: these include three negative-parity ones in the $h_{11/2}$ shell and four positive-parity ones in the $(g_{7/2}d_{5/2}d_{3/2}s_{1/2})$ shell with low $j$ values. 
One unpaired hole always occupies the $h_{11/2}$ shell and, thus, the alignment along the $z$ axis, i.e., $l$ axis, is almost constant. 
As $\omega$ becomes larger, the increase in angular momentum is generated mostly along the $x$ axis, i.e., $i$ axis for the 1-qp band and $s$ axis for the 3-qp one, through mixing of orbitals with large $j_x$ components. 
This is similar to the mechanism pointed out in the previous self-consistent investigations without pairing of Refs.~\cite{Zhao2011Phys.Rev.Lett.122501,Zhao2012Phys.Rev.C54310}. 
Pairing always reduces the alignment of the low-$j$ orbitals in the $gds$ shell, and this is connected with the fact that it reduces the collective moment of inertia by allowing partial occupation of the single-particle orbitals. 
Moreover, it is interesting to note that the alignment $J_x$ for the three $\nu(h_{11/2})$ holes rises with the inclusion of pairing, which indicates that additional admixtures are also introduced to the valence particles (holes) orbitals. 
Therefore, it is clear that the self-consistent nucleonic Cooper-pair dynamics of pairing correlations influences the single-particle orbitals and their occupation probabilities, i.e., allows additional mixing in the single-particle orbitals, and, thus, influences the generation of the nuclear spin. 
Moreover, due to the strong Coriolis term at high frequency, pairing effects become weaker and, thus, the increment of the alignment $J_x$ becomes smaller as well. 

\begin{figure}[!htbp]
\centering
\includegraphics[width=7cm]{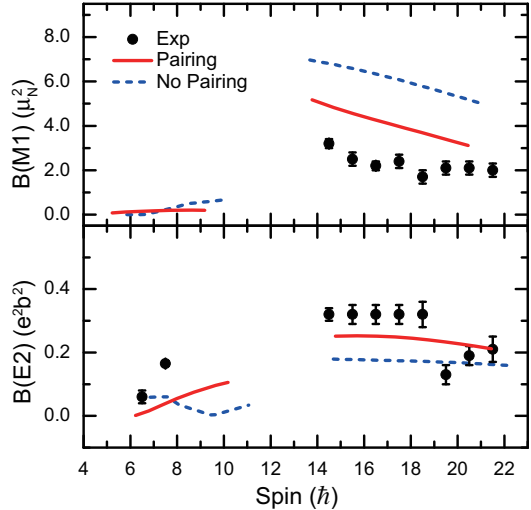}
\caption{(Color online) Calculated $B(M1)$ (upper panel) and $B(E2)$ (lower panel) values as functions of the angular momentum in comparison with data~\cite{Mukhopadhyay2007Phys.Rev.Lett.172501} (solid dots).  }
\label{fig5}
\end{figure}

Finally, the calculated transition probabilities are given in Fig.~\ref{fig5} in comparison with the available data~\cite{Mukhopadhyay2007Phys.Rev.Lett.172501}. 
Good agreement is achieved after pairing correlations are included, especially for the 3-qp band. 
The $B(M1)$ values are derived from the relativistic expression of the electromagnetic current operator, which includes both the Dirac and anomalous currents~\cite{Peng2008Phys.Rev.C24313}. 
Since the first relativistic TAC calculation~\cite{Madokoro2000Phys.Rev.C61301}, an artificial factor of 0.3 has been used for many years to attenuate the oversized values of $B(M1)$. Here, however, it is found that this factor is not needed as an increase in $B(M1)$ probabilities is counteracted mostly by  pairing, which reduces the transverse magnetic moment by merging the directions of the proton and neutron angular momenta, as shown in Fig.~\ref{fig3}. 
In addition, it should be noted that the nuclear deformation is almost unchanged by including pairing. The slight rise of the $B(E2)$ values is mainly due to the fact that pairing moves the rotational axis slightly away from the $l$ axis, i.e., there is more susceptibility to rotational alignment.

\section{Summary}
In summary, the first tilted axis cranking covariant DFT with pairing correlations has been 
formulated and the evolution of the spin axis in the yrast band of $^{135}$Nd has been investigated in a fully self-consistent microscopic way. 
The longstanding question of how pairing correlations influence the structure of tilted axis rotating nuclei has been addressed.
The present work shows that the experimental energy spectrum and the transition probabilities are well reproduced when pairing effects are taken into account.
In particular, the artificial renormalization for both the bandhead energy and the $B(M1)$ values is eliminated by including pairing. 
Moreover, it is found that the superfluidity induced by pairing allows additional mixing in single-particle orbitals, and influences the generation of the total spin, i.e., reducing the magnitude of the proton and neutron angular momenta, but expediting the merging of their directions. 
 
\begin{acknowledgments}
The authors are grateful to R. V. F. Janssens for helpful discussions and a careful reading of the manuscript.  
This work is supported by U.S. Department of Energy (DOE), Office of Science, Office of Nuclear Physics, under contract no. DE-AC02-06CH11357, by the Chinese Major State 973 Program No. 2013CB834400, and by the NSFC (Grants No. 11175002, 11105005, 111335002, 11375015, 11461141002). It used the computing resources of the Laboratory Computing Resource Center at Argonne National Laboratory.
\end{acknowledgments}

%

\end{document}